# On the Energy Analysis of Two-phase Flows Simulated with the Diffuse Interface Method

Ali Mostafavi[a], Mohammadmahdi Ranjbar[a], Vitaliy Yurkiv[a], Alexander L. Yarin[b], Farzad Mashayek[a,*]

[a] *Department of Aerospace and Mechanical Engineering, University of Arizona, Tucson, AZ 85721, USA*

[b] *Department of Mechanical and Industrial Engineering, University of Illinois Chicago, Chicago, IL 60607, USA*

The Phase-Field Method (PFM) is employed to simulate two-phase flows with the fully-coupled Cahn-Hilliard-Navier-Stokes (CHNS) equations governing the temporal evolution. The methodology minimizes the total energy functional, accounting for diffusive and viscous dissipations. A new perspective is presented by analyzing the interplay between kinetic energy, mixing energy, and viscous dissipation using the temporal evolution of the total energy functional. The classical surface energy is approximated with mixing energy under specific conditions, and the accuracy of this substitution is rigorously evaluated. The energy-based surface tension formulation derived from the Korteweg stress tensor demonstrates exceptional accuracy in capturing variations in the mixing energy. These concepts are demonstrated by considering two benchmark problems: droplet oscillation and capillary thread breakup. Key findings include validating mixing-energy theory for highly deformed interfaces, as well as the discovery of distinct energy dissipation patterns during thread breakup and droplet oscillations. The results highlight the robustness of the free energy-based PFM in accurately capturing complex interfacial dynamics, while maintaining energy conservation.

## 1. Introduction

Multiphase flow of immiscible fluids, like oil and water, which inherently resist mixing and blending, are prevalent in both nature and industry. Key applications where multiphase flows are crucial include drop impact [1–3], porous media flow [4,5], thermal management [6–8], additive manufacturing [9–12], micro- and nanofluidics [13–16], and inkjet printing [17,18]. Accurate modeling of such flows is highly sought after and has been a subject of numerous studies in the field of computational fluid dynamics (CFD).

---

The three most widely used approaches in modeling of such flows are Volume-of-fluid (VOF) [19], Level-set [20], and Phase-field methods (PFM). Typically, VOF and Level-set methodologies employ a sharp interface, while PFM adopts a diffusive interface with a finite, adjustable thickness. Classical methods for two-phase flow simulations involve interface matching conditions and surface tension localization. One natural advantage of the phase-field methods is that they do not require separate treatment at the interface. Instead, the method employs the thermodynamics-based arguments and the free energy approximation to derive the governing equations of the flow. The adoption of the PFM in two-phase flow simulations can be divided into Cahn-Hilliard [21-23], Allen-Cahn [24], and conservative second-order phase-field formulations [25].

The original fourth-order Cahn-Hilliard equation ($H^{-1}$ gradient flow method) is in a conservative format, suitable for two-phase flow simulations without phase change. The non-conservative second-order Allen-Cahn equation ($L^2$ gradient flow method) requires special attention to maintain mass conservation, such as using Lagrange multipliers [26]. More recently, the combination of the Cahn-Hilliard and Allen-Cahn models has been used to simulate three-phase flows involving icing, with the Allen-Cahn model effectively capturing the phase change [27,28].

The use of the conservative second-order phase-field method is gaining popularity [25,29,30]. The method is numerically advantageous as it yields conservative solutions, such as those governed by the Cahn-Hilliard equation, while reducing the complexity of the fourth-order derivative to a second-order derivative, as in the Allen-Cahn equation. As a result, it combines the conservative property of the Cahn-Hilliard equation with the lower derivative order of the Allen-Cahn equation, offering a more computationally efficient alternative. However, perhaps the most unique attribute of the original Cahn-Hilliard and Allen-Cahn phase-field equations is that they admit an energy law [31]. They can predict the transient behavior of multiphase media via minimization of the energy functional. This characteristic sets the Cahn-Hilliard and Allen-Cahn phase-field equations apart from other interface-capturing and interface-tracking techniques, as well as the conservative second-order phase field method, which do not adhere to a mathematically provable energy-minimization law. The dissipative energy functional, which decreases monotonically over time, provides critical insights into phase-field evolution and serves as a key indicator of numerical consistency. There have been considerable efforts in designing energy-stable numerical schemes in phase-field simulations with diffusive interface approach in the literature, often without delving deeply into their physical interpretations, especially in the fluid-flow context. This work aims to address this gap by applying the mathematical theorems developed in Ref. [32] to two-phase flow benchmark tests and revealing their physical insights.

Another advantage of the phase-field methods is the existence of a highly accurate approximation of surface tension, known as the energy-based surface tension formulation from the Korteweg stress tensor [32]. The superior performance of the energy-based surface tension model, compared to the continuum surface force model (CSF) [33] and its localized variant, is discussed in [34]. This approach reduces spurious currents and eliminates the need for direct surface curvature computation, offering an efficient and easy-to-implement method for estimating the surface tension effects. Among the other advantages, the phase-field method includes certain built-in assumptions that establish a link to classical methods for approximating surface energy. A key objective of the present study is to conduct consistency checks and determine the conditions under which these assumptions are most appropriate [35,36].

The paper is organized as follows. In section 2, the governing equations and the energy-related stability criterion are formulated. Section 3 discusses the mixing energy and the energy-based surface tension formulation. In section 4, the energy conservation equation is derived from the momentum balance equation, followed by a discussion on the work of the capillary force. Section 5 provides some information about the numerical framework and temporal and spatial discretization methods. In section 6, the suggested theories are applied to the droplet oscillations and capillary thread breakup used as benchmark problems. Conclusions are drawn in section 7, which also lists the future aims.

## 2. Governing equations

2.1. Coupled Cahn-Hilliard Navier Stokes equations

Consider a domain $\Omega \subset \mathbb{R}^D$ ($D \leq 3$) with the boundary $\Gamma$. The dynamics of the two-phase flow are governed by the coupled Cahn-Hilliard-Navier-Stokes equations, formulated in the absence of gravity, as follows:

$$\nabla \cdot \boldsymbol{u} = 0, \tag{1}$$

$$\rho \left(\frac{\partial \boldsymbol{u}}{\partial t} + \boldsymbol{u} \cdot \nabla \boldsymbol{u}\right) + (\boldsymbol{J} \cdot \nabla)\boldsymbol{u} = -\nabla P + \nabla \cdot \boldsymbol{\tau} + \boldsymbol{F}_{st}, \tag{2}$$

$$\frac{\partial c}{\partial t} + \boldsymbol{u} \cdot \nabla c = \nabla \cdot (M \nabla \psi), \tag{3}$$

$$\psi = \sigma \left(\frac{\partial f(c)}{\partial c} - \nabla^2 c\right). \tag{4}$$

Equations (1) and (2) describe the mass conservation of an incompressible medium and the momentum balance, respectively, with $\boldsymbol{u}$ being the velocity vector, $\boldsymbol{\tau} = \mu(\nabla \boldsymbol{u} + \nabla \boldsymbol{u}^T)$ the viscous stress tensor, and $P$ the fluid pressure. The density, $\rho$, and viscosity, $\mu$, of the two-phase flow are both functions of the phase-field variable in the case of unmatched densities. Typically, linear or harmonic interpolations are employed to calculate material properties inside the computational domain for the one-fluid model. It might be considered a limitation of the current phase-field formulation that the incompressibility condition ($\nabla \cdot \boldsymbol{u} =$

0) is enforced as the continuity equation, despite the occurrence of density variation at the interface. Yue [37] considered the effects of compressibility and found that they are confined to the interfacial region, with minimal impact on the macroscopic flow in droplets with contact angle hysteresis oscillating on a solid surface. A perfect solenoidal velocity is not expected in the case of unmatched densities ($\nabla \cdot \boldsymbol{u} \neq 0$). However, providing sufficient mesh elements to decrease the density variation across the elements in the interface region can mitigate the issue. This is because equation (1) holds strictly for single-phase flows with constant density. Across the diffusive interface, smooth transition of material properties leads to a non-divergence-free velocity field. In addition, $\boldsymbol{F}_{st}$ is the energy-based capillary force, which will be derived later in the Appendix. Also, $\boldsymbol{J} = \frac{\rho_2 - \rho_1}{2} M \nabla \psi$ represents the diffusive flux arising from the disparities between mass- and volume-averaged velocities and is proportional to the density difference between phases [32]. Although $(\boldsymbol{J} \cdot \nabla)\boldsymbol{u}$ is typically small and often neglected in engineering applications, it is retained in the momentum balance equation to ensure thermodynamic consistency, as highlighted in [32].

Equations (3) and (4) denote the split form of the advective Cahn-Hilliard phase-field equation, where $c$ is the order parameter, $\psi$ is the chemical potential, $M$ is the phenomenological mobility coefficient, $f(c)$ is the bulk free energy, and $\sigma$ is the mixing energy density. A global labeling function $c$ is defined to differentiate between separate phases: $c = 1$ represents fluid 1, $c = -1$ indicates fluid 2, $-1 < c < 1$ corresponds to the diffusive interface layer, with $c = 0$ reflecting the fluid–fluid sharp interface.

The mixing energy per unit volume is given by

$$f_m(c, \nabla c) = \sigma \left( \frac{1}{2} |\nabla c|^2 + f(c) \right), \tag{5}$$

where,

$$f(c) = \frac{(c^2 - 1)^2}{4\xi^2}, \tag{6}$$

is the double-well potential, with $\xi$ being the capillary width, also known as the interface thickness. This equation recognizes a phobic bulk component, $\sigma f(c)$, which expresses the tendency for phase separation, and a philic surface component, $\frac{\sigma}{2}|\nabla c|^2$, which expresses the mixing tendency [35].

The original Cahn-Hilliard equation governs the evolution of the conserved variable $c$ by facilitating the dissipation of the free energy functional,

$$\mathcal{F} = \int_\Omega \bigl(f_m(c, \boldsymbol{\nabla} c)\bigr) d\Omega. \tag{7}$$

The free energy functional $\mathcal{F}$ (also known as the Ginzburg-Landau free energy) represents the total free energy over the domain $\Omega$. It consists of the local free energy density $f(c)$, which represents the bulk free energy and the gradient energy density $f_{gr} = \frac{\sigma}{2}|\boldsymbol{\nabla} c|^2$, both of which act at the diffusive interface. Hence, the total mixing energy, when no additional source of energy is present, is $\mathcal{F}$. Commonly, the double-well free function is used to delineate the local free energy $f(c)$ in two-phase flow simulations [Eq. (6)]. The chemical potential $\psi$ is defined as the variational derivative of the total free energy $\mathcal{F}$ with respect to the phase-field variable $c$, i.e., $\psi = \delta\mathcal{F}/\delta c$. This formulation ensures that in the absence of fluid flow, the multiphase media reach an equilibrium state by minimizing their free energy, as the general thermodynamics principles imply.

2.2. Energy stability

The phase-field evolution follows a gradient descent in the energy space, leading to a dissipative Cahn-Hilliard or Allen-Cahn equation. In multiphase fluid flows, additional forms of energy are involved, such as the kinetic energy and potential energy of the flow. The link between the momentum balance and the phase-field equations is established implicitly by taking the inner product of Eq. (3) with $-\psi$, Eq. (4) with $\partial c/\partial t$ and Eq. (2) with $\boldsymbol{u}$, and then summing the relations, obtaining the following dissipative energy law under zero-gravity condition [35,38,39]:

$$\frac{dE_{tot}}{dt} = -\int_\Omega \left(\frac{\mu}{2}\|\boldsymbol{\nabla}\boldsymbol{u}\|_F^2 + M\|\boldsymbol{\nabla}\psi\|^2\right) d\Omega, \tag{8}$$

where,

$$E_{tot} = \int_\Omega \left(\frac{1}{2}\rho\|\boldsymbol{u}\|^2 + f_m(c, \boldsymbol{\nabla} c)\right) d\Omega = E_k + E_m. \tag{9}$$

The first term on the right-hand side of Eq. (8) is viscous dissipation, whereas the second one corresponds to the chemical dissipation or diffusive dissipation. The norms designated in Eqs. (8) and (9) are the Frobenius matrix norm and the Euclidean vector norm. These norms are, respectively, defined as follows [39]:

$$\|\nabla \boldsymbol{u}\|_F^2 = \sum_i \sum_j \left|\frac{\partial u_i}{\partial x_j}\right|^2, \qquad \|\boldsymbol{u}\|^2 = \sum_i |u_i|^2. \tag{10}$$

Equation (8) expresses that the total energy $E_{tot}$, comprising kinetic energy $E_k$ and mixing energy $E_m$, is irreversibly dissipated through both viscous and diffusive energy loss, with the latter occurring via dissipative diffusion driven by the chemical potential gradient. The Cahn-Hilliard two-phase flow simulations uniquely minimize $E_{tot}$, setting them apart from the methods like VOF, Level-set, and the second-order phase-field, which lack a known energy law [34]. A part of the phase-field literature has been devoted to developing numerical schemes possessing energy stability, meaning they exhibit dissipative behavior over time. Various numerical schemes such as convex splitting [40], linear stabilization [41], invariant energy quadratization [42] and scalar auxiliary variable [43] have been developed [37]. The energy stability feature must be preserved in different numerical schemes unconditionally. Since the main purpose of the present work is to offer new perspectives on studying two-phase flow dynamics from an energy standpoint, the time derivative of the total energy, $\dot{E}_{tot}$, is also monitored, as it provides significant insight in the context of fluid flow. It directly relates viscous and diffusive dissipations to the energy functional and highlights the stages at which the flow is predominantly dissipative.

### 3. Mixing energy and capillary force

Consider a $1D$ case, where an interface is at equilibrium without fluid flow ($\boldsymbol{u} = 0$), the diffusive flux must vanish at the interface, i.e., $\psi = \delta \mathcal{F}/\delta c = 0$ [35,44]. Equation (11) describes the equilibrium state of the conserved order parameter, as derived from Eq. (4):

$$\sigma \left( -\frac{d^2 c}{dx^2} + \frac{\partial f(c)}{\partial c} \right) = 0, \tag{11}$$

where substituting $f(c)$ from Eq. (6) leads to the solution

$$c(x) = \tanh\left(\frac{x}{\sqrt{2}\xi}\right). \tag{12}$$

For a detailed discussion of the boundary conditions associated with Eq. (11), see [35]. In this scenario, the gradient energy and the derivative of the bulk energy offset one another, yielding the equilibrium condition specified in Eq. (12). The distributions of the order parameter, $c$, and the bulk free energy density, $f(c)$, for various interfacial thicknesses at the equilibrium state are given in Fig. 1 (a) and Fig. 1 (b), respectively, for a computational domain size of $\Delta x$.

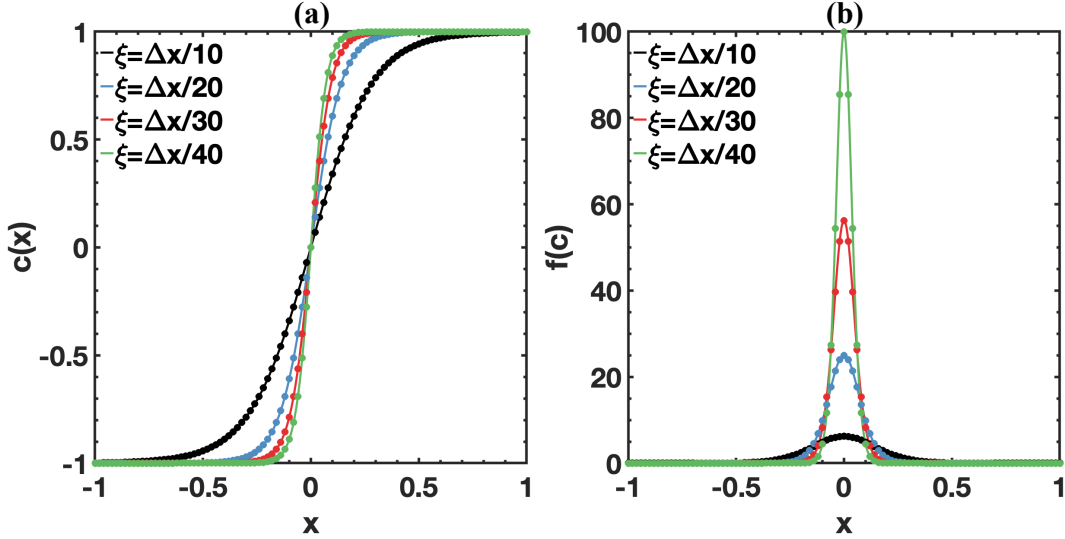

**Fig. 1.** Results for a $1D$ interface at equilibrium without fluid flow. Displayed in the panels are (a) the equilibrium profiles of the order parameter and (b) the bulk free energies for various values of the interfacial thickness.

The relation between the diffuse interface approach and the classical concept of the interfacial tension is established by equating the total free energy with the traditional surface energy, as follows [35]:

$$\gamma = \sigma \int_{-\infty}^{\infty} \left\{ \frac{1}{2}\left(\frac{dc}{dx}\right)^2 + f(c) \right\} dx. \tag{13}$$

Upon substituting the equilibrium profile $c(x) = \tanh(x/\sqrt{2}\xi)$ for the order parameter into Eq. (13) and performing integration, the following relationship expresses the relation between the mixing energy density $\sigma$ and the interfacial surface tension coefficient $\gamma$:

$$\sigma = \frac{3}{2\sqrt{2}} \gamma \xi. \tag{14}$$

Equation (14) yields exact results regardless of the interfacial thickness. Despite this, the relation is derived under specific assumptions— a $1D$ condition at the equilibrium without fluid flow, and the absence of the gradient in the chemical potential. In practical applications, one seeks to accurately approximate the surface energy using the mixing energy theory in transient simulations where the chemical potential gradient remains non-vanishing. With phase-field modeling, it is still required to use Eq. (14) to approximate the surface energy. Another issue that compromises the underlying assumptions in Eq. (14) is the unboundedness of the solution of the Cahn-Hilliard equation arising from the biharmonic operator [45,46]. This causes the order parameter $c$ to be prone to undershoots and overshoots, further challenging the equivalence between mixing and surface energies. It is of broad interest to examine how well Eq. (14) holds in simulations of the transient behavior of two-phase flows with curved interfaces and whether it can approximate the classical surface energy within the framework of the free energy approach. To the best of our knowledge, such investigations have not been conducted in the context of the phase-field two-phase flow simulations.

In the Cahn-Hilliard models, consistent with the second law of thermodynamics, the divergence of the Korteweg stress tensor ($\nabla \cdot (\nabla c \otimes \nabla c)$) is widely used to model capillary forces. Various formulations of the volumetric surface tension force exist in the literature, with more details available in [47]. In this work, the following expression is adopted to model the capillary force (see Appendix for a detailed derivation):

$$F_{st} = \psi \nabla c, \tag{15}$$

which leads to the following modified pressure expression:

$$p = P + \sigma \left( \frac{1}{2} \nabla c \cdot \nabla c + f(c) \right). \tag{16}$$

### 4. Energy conservation and work of the capillary force

Another way to analyze two-phase flows from the energy perspective is via taking the dot product of the momentum balance equation with the velocity vector $u$, yielding

$$\rho u \cdot \left( \frac{\partial u}{\partial t} + u \cdot \nabla u \right) + u \cdot (J \cdot \nabla) u = -u \cdot \nabla p + u \cdot (\nabla \cdot \tau) + u \cdot (F_{st}). \tag{17}$$

Here $p$ corresponds to the modified pressure, which arises directly from the term providing the surface tension expression (cf. Eq. (A6) in Appendix for the derivation of its relation to $P$). Taking the surface/volume integral of Eq. (17) leads to:

$$\int_\Omega \left( \rho u \cdot \frac{\partial u}{\partial t} + \rho u \cdot (u \cdot \nabla u) + u \cdot (J \cdot \nabla) u \right) d\Omega = \int_\Omega \left( -u \cdot \nabla p + u \cdot (\nabla \cdot \tau) + u \cdot F_{st} \right) d\Omega. \tag{18}$$

Equation (18) reformulates the momentum balance, originally in vector form, into a scalar expression referred to as energy/momentum conservation. Verifying the equivalence of the left- and right-hand sides of Eq. (18) is essential for assessing the accuracy of the numerical scheme. As numerical methods inherently introduce artificial diffusion and dispersion, evaluating the energy balance provides a quantitative measure of their spurious effects. See [29,48] for investigations into the role of uniform mesh size in improving momentum conservation in interfacial flows.

For simplicity, consider first two-phase flows with matched densities and viscosities and neglect the body force (e.g., gravity). In this matched-density case, the diffusive flux is zero ($\boldsymbol{J} = \boldsymbol{0}$). Under divergence-free velocity field assumption ($\boldsymbol{\nabla} \cdot \boldsymbol{u} = 0$), the pressure gradient and the advection terms can be rewritten as,

$$\boldsymbol{u} \cdot \boldsymbol{\nabla} p = \boldsymbol{\nabla} \cdot (p\boldsymbol{u}), \qquad \rho \boldsymbol{u} \cdot (\boldsymbol{u} \cdot \boldsymbol{\nabla} \boldsymbol{u}) = \frac{1}{2} \boldsymbol{\nabla} \cdot (\rho \boldsymbol{u} |\boldsymbol{u}|^2). \tag{19}$$

Theoretically, the advection term $\int_\Omega \frac{1}{2} \boldsymbol{\nabla} \cdot (\rho \boldsymbol{u} |\boldsymbol{u}|^2)\, d\Omega$ and the modified pressure-gradient term $\int_\Omega \boldsymbol{\nabla} \cdot (p\boldsymbol{u})\, d\Omega$ should yield zero under specific conditions. This can be shown with the application of the divergence theorem once the velocity field is supplemented with periodic boundary conditions or the following boundary conditions on the domain boundaries $\Gamma$:

$$\boldsymbol{u}|_\Gamma = 0, \qquad \frac{\partial \boldsymbol{u}}{\partial \boldsymbol{n}}\bigg|_\Gamma = 0, \tag{20}$$

where $\boldsymbol{n}$ is the unit normal vector to $\Gamma$. Also, the first term on the left-hand side of Eq. (18) is the time derivative of the kinetic energy $E_k$, which will be referred to as $\partial E_k / \partial t$ here and hereinafter. In the matched-density case, $\rho$ is constant in space and time. Therefore,

$$\frac{\partial E_k}{\partial t} = \partial_t \left( \rho, \frac{|\boldsymbol{u}|^2}{2} \right) = (\rho \boldsymbol{u}_t, \boldsymbol{u}), \tag{21}$$

where $(\cdot,\cdot)$ indicates the inner product in $L^2(\Omega)$. Hence, Eq. (18) reduces to:

$$\frac{\partial E_k}{\partial t} = \int_\Omega (\boldsymbol{u} \cdot (\boldsymbol{\nabla} \cdot \boldsymbol{\tau}) + \boldsymbol{u} \cdot \boldsymbol{F}_{st})\, d\Omega. \tag{22}$$

The work of viscous forces is denoted as $\Phi_v = \boldsymbol{u} \cdot (\nabla \cdot \boldsymbol{\tau})$. For completeness, the expressions for $\Phi_v$ are provided in both $xy$ (Cartesian) and $rz$ (axisymmetric) coordinates:

$$\Phi_{v,xy} = \mu\left(u_x\left(\frac{\partial^2 u_x}{\partial x^2} + \frac{\partial^2 u_x}{\partial y^2}\right) + u_y\left(\frac{\partial^2 u_y}{\partial x^2} + \frac{\partial^2 u_y}{\partial y^2}\right)\right), \tag{23}$$

$$\Phi_{v,rz} = \mu\left(u_r\left(\frac{\partial^2 u_r}{\partial r^2} + \frac{\partial^2 u_r}{\partial z^2} + \frac{1}{r}\frac{\partial u_r}{\partial r} - \frac{u_r}{r^2}\right) + u_z\left(\frac{\partial^2 u_z}{\partial r^2} + \frac{\partial^2 u_z}{\partial z^2} + \frac{1}{r}\frac{\partial u_z}{\partial r}\right)\right). \tag{24}$$

Due to the presence of the second-order derivatives of the fluid velocity, first-order Lagrange elements can no longer be used for velocity approximation as they return zero for the second-order derivatives unless the solution is accurately reconstructed with higher-order shape functions, or the derivative's order is reduced via the divergence theorem. Alternatively, the velocity could be discretized using at least a second-order approximation to accurately compute the viscous power in the formulated approach. The schematic of a finite element utilized in this work is illustrated in Fig. 2. The Taylor-Hood pair of finite elements is used, where linear shape functions are used for $p$, $c$ and $\psi$, while quadratic shape functions are used for $\boldsymbol{u}$ to directly compute $\Phi_v$ [49,50].

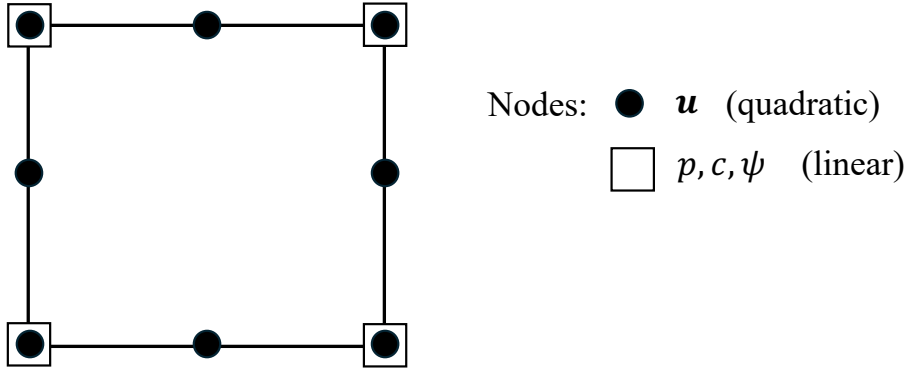

**Fig. 2.** A diagram of 2D Quad8 libMesh Lagrange element [51] used for spatial discretization. Second-order shape functions are utilized for the velocity approximation, while other nonlinear variables being interpolated using linear shape functions.

The viscous power density can be written as

$$\boldsymbol{u} \cdot (\nabla \cdot \boldsymbol{\tau}) = \nabla \cdot (\boldsymbol{\tau u}) - \boldsymbol{\tau} : \nabla \boldsymbol{u}, \tag{25}$$

where

$$\phi_v = -\boldsymbol{\tau} : \boldsymbol{\nabla} \boldsymbol{u} \leq 0, \tag{26}$$

is the local viscous dissipation [52]. Complete expressions for $\phi_v$ in various coordinate systems are given in [53]. The application of the divergence theorem entails that the total power of viscous forces is equivalent to the viscous dissipation:

$$\int_\Omega \Phi_v \, d\Omega = \int_\Omega \phi_v \, d\Omega. \tag{27}$$

Spatial integration can produce numerical errors. The most accurate way for performing such integration in Eq. (18) is to use Gaussian quadrature points rather than nodal points, which is the standard in the finite element methods. Temporal integration is carried out using the trapezoidal rule, with sufficiently small time step sizes chosen to minimize the time integration errors. Subsequently, integrating Eq. (22) over time yields:

$$E_{k,t} - E_{k,0} = \int_0^t \int_\Omega \Phi_v \, d\Omega dt + \int_0^t \int_\Omega \boldsymbol{u} \cdot \boldsymbol{F}_{st} \, d\Omega dt, \tag{28}$$

which relates the change of the kinetic energy to the work of viscous stresses ($E_v$) and capillary forces. Naturally, the change in the kinetic energy is related to the change of the surface energy and viscous dissipation. Therefore, it is anticipated that the work of the surface tension represents the change in the surface energy (essentially, a potential energy) of the flow. It is important to compare the work of capillary force $\int_0^t \int_\Omega \boldsymbol{u} \cdot \boldsymbol{F}_{st} \, d\Omega dt$ with the variation of the free energy functional/mixing energy $\int_\Omega f_m(c, \boldsymbol{\nabla} c) d\Omega$ and surface energy starting from the initial stage. For a given interface, the surface energy is equal to the product of the surface tension coefficient and the surface area. Ideally, the work of capillary forces should exactly match the changes in the mixing and surface energies; however, the derivation in Eq. (14) is valid only under a specific condition.

The extension of the arguments made thus far to two-phase flows with variable density and viscosity is somewhat different. Considering Eq. (19), the pressure-gradient term can still be incorporated into the $\boldsymbol{\nabla}$ operator, converted into a boundary integral using the divergence theorem, and subsequently eliminated. The main difference is that $\partial E_k / \partial t$ contains time derivative of density which is non-zero for flows with unmatched densities. By using linear interpolation for $\rho$ and defining $\boldsymbol{J}$ as follows,

$$\rho(c) = \frac{\rho_1 - \rho_2}{2} c + \frac{\rho_1 + \rho_2}{2}, \quad \boldsymbol{J} = \frac{\rho_2 - \rho_1}{2} M \boldsymbol{\nabla} \psi, \tag{29}$$

an equation for the temporal change of density is obtained from Eqs. (1), (3), (29), as [38]:

$$\frac{\partial \rho}{\partial t} + \boldsymbol{\nabla} \cdot (\rho \boldsymbol{u}) + \boldsymbol{\nabla} \cdot \boldsymbol{J} = 0. \tag{30}$$

Abels et al. demonstrated that the individual masses of the distinct phases remain conserved with this formulation [32]. In the case of two-phase flows with unmatched densities, $\partial E_k / \partial t$ is calculated as [38]:

$$\begin{aligned} \partial_t \left( \rho, \frac{|\boldsymbol{u}|^2}{2} \right) &= (\rho \boldsymbol{u}_t, \boldsymbol{u}) + \left( \rho_t, \frac{|\boldsymbol{u}|^2}{2} \right) = (\rho \boldsymbol{u}_t, \boldsymbol{u}) - \left( \boldsymbol{\nabla} \cdot (\rho \boldsymbol{u}) + \boldsymbol{\nabla} \cdot \boldsymbol{J}, \frac{|\boldsymbol{u}|^2}{2} \right) \\ &= (\rho \boldsymbol{u}_t + \rho \boldsymbol{u} \cdot \boldsymbol{\nabla} \boldsymbol{u} + \boldsymbol{J} \cdot \boldsymbol{\nabla} \boldsymbol{u}, \boldsymbol{u}). \end{aligned} \tag{31}$$

Equation (31) relates the temporal change in the kinetic energy to the left-hand side of Eq. (18). The term $(\rho \boldsymbol{u}_t, \boldsymbol{u})$ represents the change in the kinetic energy due to the variation in velocity, while the term $(\rho \boldsymbol{u} \cdot \boldsymbol{\nabla} \boldsymbol{u} + \boldsymbol{J} \cdot \boldsymbol{\nabla} \boldsymbol{u}, \boldsymbol{u})$ accounts for the change in the kinetic energy due to the variation in density. Although due to the density variation the expression for $\partial E_k / \partial t$ is different for the matched and unmatched cases, Eq. (28) is applicable in both cases. The energy analysis has been extensively performed for single-phase flows [52,54], but its application to two-phase flow benchmark tests within the framework of the Cahn-Hilliard-Navier-Stokes (CHNS) equations is lacking, despite the existence of well-established general theories [32,38]. In the diffuse-interface picture, two-phase flows are treated as a single-fluid model, and the conservation laws are solved in the two-phase flow domain. Omitting the term $(\boldsymbol{J} \cdot \boldsymbol{\nabla})\boldsymbol{u}$ from the momentum balance equation invalidates Eq. (31) and introduces thermodynamic inconsistency.

In summary, this section introduces several approaches to analyzing two-phase flows from an energy perspective. Additionally, some methods for assessing the numerical accuracy and consistency of the corresponding solutions are discussed. The next section outlines the details of the computational framework, followed by the discussion of the results in section 6.

## 5. Numerical methods

Equations (1)-(4) are solved by the finite element method using the continuous Galerkin discretization within the MOOSE framework, employing the Taylor-Hood pair of finite element spaces [55]. More details on the two-phase flow solver in the MOOSE framework, including its numerical implementation and validation, are available elsewhere [46]. The first-order $Q1$ elements are used to discretize $p$, $c$, and $\psi$, while the second-order $Q2$ elements are used for $\boldsymbol{u}$. Furthermore, the use of the second-order velocity approximation results in a better conservation of the momentum equation compared to the first-order approximation [56]. The full weak form of the governing equations can be expressed as:

$$\left(\rho\frac{D\boldsymbol{u}}{Dt} + \boldsymbol{J}\cdot\boldsymbol{\nabla}\boldsymbol{u},\hat{\boldsymbol{u}}\right)_\Omega - (\psi\boldsymbol{\nabla}c,\hat{\boldsymbol{u}})_\Omega + (-p\boldsymbol{I}+\boldsymbol{\tau},\boldsymbol{\nabla}\hat{\boldsymbol{u}})_\Omega - (\boldsymbol{n}\cdot(-p\boldsymbol{I}+\boldsymbol{\tau}),\hat{\boldsymbol{u}})_\Gamma = 0, \tag{32}$$

$$-(\boldsymbol{\nabla}\cdot\boldsymbol{u},\hat{p})_\Omega = 0, \tag{33}$$

$$\left(\psi - \sigma\frac{\partial f}{\partial c},\hat{c}\right)_\Omega - (\sigma\boldsymbol{\nabla}c,\boldsymbol{\nabla}\hat{c})_\Omega + (\boldsymbol{n}\cdot(\sigma\boldsymbol{\nabla}c),\hat{c})_\Gamma = 0, \tag{34}$$

$$\left(\frac{\partial c}{\partial t},\hat{\psi}\right)_\Omega + (\boldsymbol{u}\cdot\boldsymbol{\nabla}c,\hat{\psi})_\Omega + (M\boldsymbol{\nabla}\psi,\boldsymbol{\nabla}\hat{\psi})_\Omega - (\boldsymbol{n}\cdot(M\boldsymbol{\nabla}\psi),\hat{\psi})_\Gamma = 0. \tag{35}$$

Variables marked with the symbol ^ are the corresponding test functions, and $D/Dt = \partial/\partial t + \boldsymbol{u}\cdot\boldsymbol{\nabla}$ is the material derivative operator. The governing equations are solved in a fully-coupled and fully-implicit manner altogether. The temporal discretization is performed using a finite difference approximation, with time integration implemented through the second-order Backward Differentiation Formula (BDF2), providing second-order accuracy in time. The nonlinear equations are solved using the Newton method [57]. At each Newton iteration, the linearized system is solved using the MUltifrontal Massively Parallel Sparse direct Solver (MUMPS), which efficiently handles large sparse systems via LU factorization.

The simulations in the present work significantly benefit from the application of the adaptive mesh refinement (AMR) at the interface. Each $D$-dimensional element is refined isotropically by subdividing it into $2^D$ geometrically similar child elements. The 1 : 2 refinement ratio between successive levels ensures that each child element inherits the aspect ratio of its parent. When necessary, coarsening simply reverses this subdivision. Mesh adaptation within the computational domain is guided by a gradient jump indicator, evaluated at the quadrature points as follows:

$$Jump = \left(((\boldsymbol{\nabla}c)_k - (\boldsymbol{\nabla}c)_{K'})\cdot\boldsymbol{n}\right)^2, \tag{36}$$

where $(\boldsymbol{\nabla}c)_K$ and $(\boldsymbol{\nabla}c)_{K'}$ are the phase-field gradients on element $K$ and its adjacent neighbor $K'$, respectively, and $\boldsymbol{n}$ is the outward unit normal to their common face. The jump quantifies the interface sharpness and precisely localizes mesh refinement to regions of steep phase transition.

Although AMR provides a sufficient resolution and effectively models thin interfaces, the interpolation error introduced during the mesh coarsening is a drawback [39,46]. In our previous work, a Lagrange multiplier was introduced to reduce the mesh-coarsening effects on the global conservation of the phase-field variable within the machine precision error with accurate physical outcomes [46]. The simulations leverage parallel computing capabilities by utilizing the METIS package for efficient domain partitioning, which dynamically optimizes the workload distribution across multiple processors each time step [58]. For an in-depth discussion of dynamic domain partitioning in two-phase flow simulations, see [46]. An illustration of the computational mesh, leveraging AMR is presented in Fig. 3.

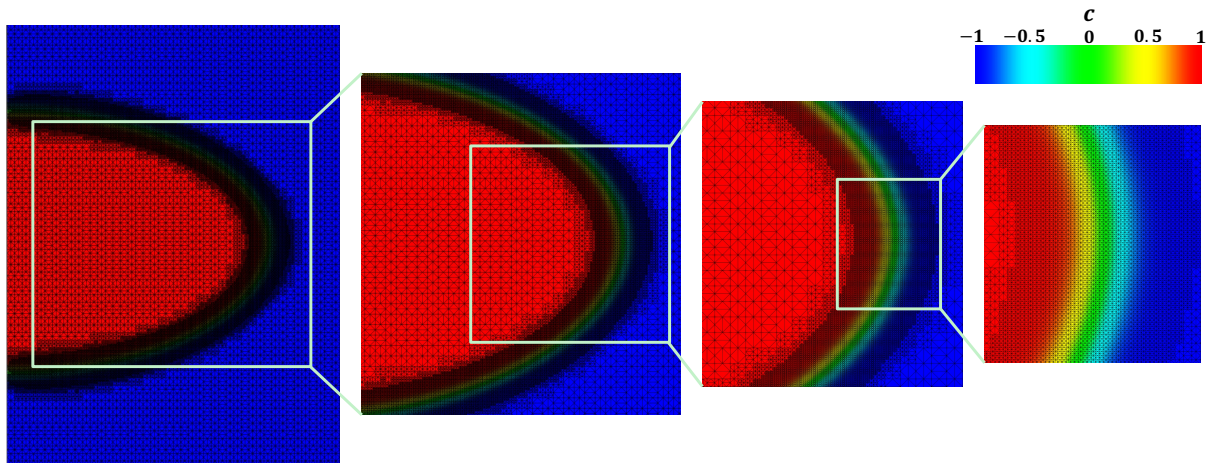

**Fig. 3**. Representation of AMR at the interface. The figures show the computational domain adjacent to the diffusive interface, discretized by a $220 \times 220$ base mesh with three levels of adaptive mesh refinement. The zoomed-in insets offer magnified views of the mesh at the diffusive interface, with approximately 15 elements resolving the thin interface.

## 6. Results and discussion

6.1. Static tests for surface energy approximation

While Eq. (14) yields exact results for approximating surface energy under the mixing-energy theory in the one-dimensional case, its extension to more general scenarios remains subject to verification. This includes cases such as curved interfaces under stationary conditions without fluid flow. Therefore, static tests are conducted here to evaluate the accuracy of Eq. (14) in approximating the surface energy of axisymmetric interfaces initialized from non-equilibrium configurations. The analysis considers initially oblate spheroidal droplets subjected to various degrees of deformation. Given that the mixing energy density $\sigma$ is derived under the assumption of a hyperbolic tangent profile for the order parameter $c$, it is critical to initialize the diffuse interface such that the transition from $c = 1$ to $c = -1$ adheres strictly to a hyperbolic tangent distribution, as depicted in Fig. 1 (a). Any alternative initialization deviating from the tanh form compromises the theoretical foundation of Eq. (14) and therefore necessitates a reformulation of $\sigma$.

The signed distance function $d(r,z)$ facilitates initialization of the phase-field across the diffusive interface [59]. It measures the signed distance from each nodal point to the sharp ellipsoidal interface—positive outside, negative inside, and zero on the interface—ensuring a consistent transition of the order parameter. For ellipsoidal geometries, $d(r,z)$ is computed as the signed Euclidean distance from each nodal point to the $c = 0$ sharp-interface contour. The order parameter is initialized using the standard tanh profile from the steady-state solution of the one-dimensional phase-field model while ensuring that the interface thickness remains uniform in the direction normal to the interface:

$$c(r,z) = \tanh\left(-\frac{d(r,z)}{\sqrt{2}\xi}\right), \qquad (37)$$

where $\xi$ controls the thickness of the transition layer.

Panel (a) in Fig. 4 illustrates the consistent initialization of the order parameter based on Eq. (37), whereas panel (b) in the same figure shows the distribution of the order parameter along directions normal to the interface at various angular positions for a spheroidal droplet with an initial oblate aspect ratio of 2. In these figures, $r^*$ and $z^*$ represent non-dimensional lengths normalized by the droplet's volume-equivalent radius $R$. Panel (c) displays oblate droplet isocontours at $c = -0.9, 0.0, 0.9$, demonstrating a uniform interface width measured normal to the interface. The derivation of Eqs. (13) and (14) relies on a constant interface thickness measured normal to the interface—not, for example, in the radial direction. Panel (d) indicates the normalized curvature distribution along the ellipsoidal interface for aspect ratios ranging from 1 to 4, computed as

$$\kappa = -\nabla \cdot \left(\frac{\nabla c}{|\nabla c|}\right). \qquad (38)$$

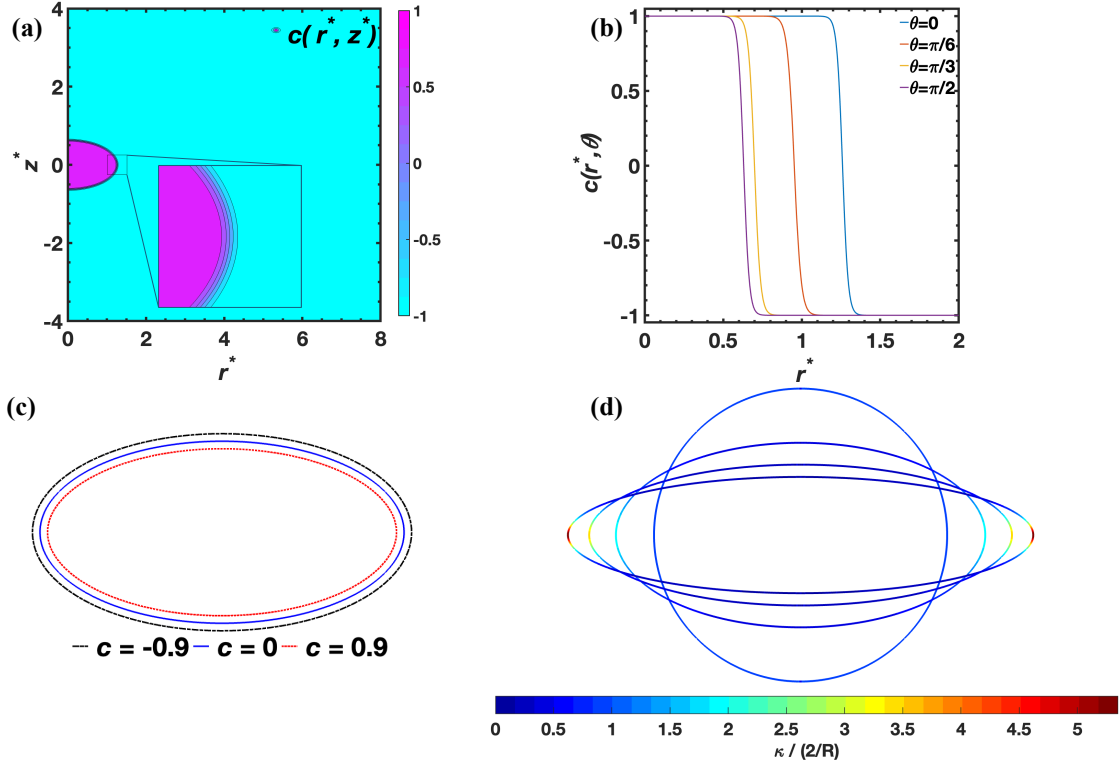

**Fig. 4.** (a) Consistent initialization of the order parameter using a hyperbolic-tangent profile derived from a signed-distance function in an axisymmetric domain; (b) radial cuts of the normalized order parameter normal to the interface at various polar angles $\theta$ (in radians); (c) oblate-droplet isocontours at $c = -0.9$, 0.0 and 0.9, showing uniform interface thickness normal to the interface; (d) spatial heterogeneity of normalized curvature along the oblate-droplet interface for aspect ratios ranging from 1 to 4.

Consequently, static tests are performed on droplets with curved interfaces of varying curvature to assess the validity of the mixing energy theory under different mesh resolutions and interfacial thicknesses. Unlike Fig. 4 (a), which uses non-dimensional coordinates, a dimensional axisymmetric computational domain of $[0, 8R] \times [-4R, 4R]$ is considered for next simulations. Spheroidal droplet shapes with various prescribed oblate aspect ratios are considered. For interfacial profiles with axial symmetry, the surface energy ($E_s$) is calculated by the product of the surface area with the surface tension coefficient:

$$E_s = 2\pi\gamma \int_{z_1}^{z_2} r(z) \sqrt{1 + \left(\frac{dr}{dz}\right)^2} \, dz, \tag{39}$$

where $r(z)$ is extracted as the isocontour of $c = 0$, representing the sharp fluid–fluid interface. Thus, for a given order parameter distribution $c$, Eq. (7) yields the total mixing energy $E_m$, while Eq. (39) determines the surface energy $E_s$. The comparison between these forms of energy is presented in Fig. 5.

In the static tests, a uniform grid is employed to initialize the order parameter and to perform the numerical integration using a four-point Gaussian quadrature rule. The results in Fig. 5 (a) assume an interfacial thickness of $\xi = R/20$, where $n$ denotes the number of uniform divisions in both radial and axial directions. Across all aspect ratios, these plots reveal that insufficient mesh resolution underestimates the mixing energy relative to the surface energy. Once the diffusive interface is sufficiently resolved, mixing-energy theory accurately recovers the surface energy for both spherical droplet of uniform curvature and highly deformed droplets. Additionally, the plots in Fig. 5 (b) investigate the influence of the interfacial thickness on the surface-energy approximation using a $2000 \times 2000$ uniform grid. These results indicate that, for a fixed mesh size across all capillary widths, decreasing the interfacial thickness reduces the discrepancy between the surface energy and the mixing energy. This observation aligns with the well-known premise that Cahn-Hilliard models increasingly approximate the sharp-interface limit as the interface thickness is reduced below a critical threshold [60,61].

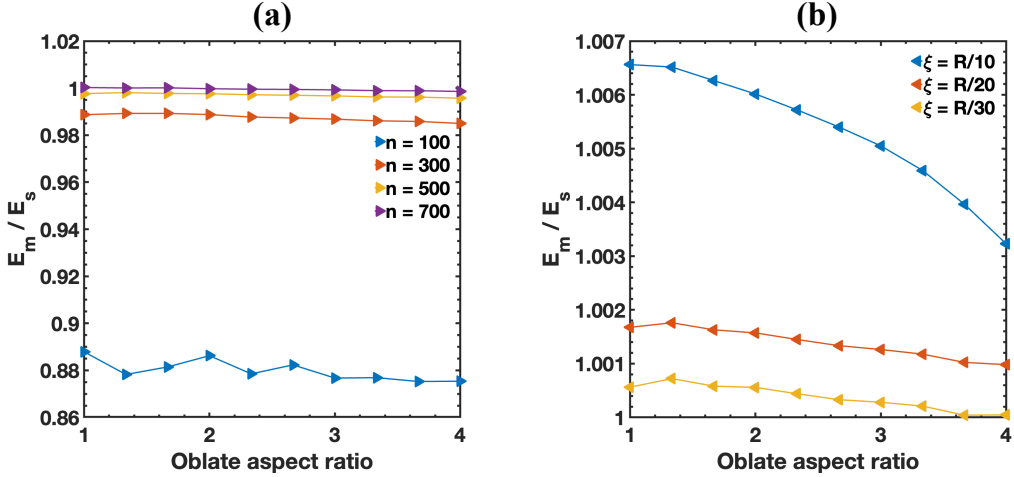

**Fig. 5.** Comparison of the mixing energy and surface energy for a curved interface. Panel (a) shows the effect of grid resolution, and panel (b) illustrates the effect of interfacial thickness on the surface energy approximation, each plotted as a function of the initial perturbation.

Overall, the static tests confirm that Eq. (14) accurately recovers surface energy in phase-field simulations when the interface is resolved with a sufficient number of elements. This validity extends to curved interfaces when the phase-field variable is initialized consistently. In particular, the transition from $c = -1$ to $c = 1$ must strictly follow the hyperbolic-tangent profile, and the interface thickness must remain constant along the direction normal to the interface.

### 6.2. Matched-density droplet oscillation

The oscillation dynamics of an initially ellipsoidal droplet in a two-phase medium is used as a benchmark problem for the energy analysis. The arguments are extendable to many types of two-phase flows with the same underlying physics. A schematic of the non-dimensional computational domain is shown in Fig. 4 (a), while Fig. 3 displays the computational grid near the interface. Henceforth, all simulations are conducted on adaptive grids using adaptive mesh refinement (AMR). Large-amplitude oscillations of an initially oblate spheroidal droplet with an initial aspect ratio of 2 are simulated. The analysis starts with matched density and viscosity and is later expanded to different material properties. In Fig. 4 (a), the $z$-axis is the symmetry axis, and all variables are supplemented with zero-flux boundary conditions, which does not require special treatment in finite element method, except for $u_r = 0$ at the symmetry axis. This set of boundary conditions theoretically zeroes out the integrals $\int_\Omega -\boldsymbol{u} \cdot \nabla p \, d\Omega$ and $\int_\Omega \rho \boldsymbol{u} \cdot (\boldsymbol{u} \cdot \nabla \boldsymbol{u}) \, d\Omega$.

The fluid parameter values are assumed to be $\rho = 10^3$ kg/m³, $\mu = 10^{-3}$ Pas and $\gamma = 0.067$ N/m [62]. The oscillation dynamics is fully characterized by the Reynolds number ($Re = \rho U R / \mu$) and the Weber number ($We = \rho U^2 R / \gamma$) in the absence of gravity and external forces. For the case of matched density and viscosity, the Reynolds number is set to $Re = 200$. The velocity scale $U$ is obtained by setting $We = 1$. Then, the radius of the unperturbed spherical droplet, which also serves as the characteristic length, is given by:

$$\tilde{R} = \frac{Re^2 \mu^2}{We \rho \gamma}. \tag{40}$$

The characteristic velocity, time, energy, and the temporal rate of energy variation are defined, respectively, as:

$$\tilde{U} = \frac{Re \mu}{\rho \tilde{R}}, \quad \tilde{t} = \frac{\tilde{R}}{\tilde{U}}, \quad \tilde{E} = \frac{1}{2}\rho \tilde{U}^2 \mathcal{V}, \quad \tilde{\dot{E}} = \frac{\tilde{E}}{\tilde{t}}, \tag{41}$$

where $\mathcal{V}$ is the volume of a spherical droplet. These characteristic numbers are used to render the results dimensionless, although the governing equations are solved in a dimensional form.

The mobility parameter remains constant, and the interfacial thickness is initially chosen as $\xi = R/20$. In Fig. 6 (a), the volume integration in Eq. (18) is performed at the quadrature points. The plots show the dimensionless volume integrals of $\rho \boldsymbol{u} \cdot \partial \boldsymbol{u}/\partial t$, $\rho \boldsymbol{u} \cdot (\boldsymbol{u} \cdot \nabla \boldsymbol{u})$, $-\boldsymbol{u} \cdot \nabla p$, $\Phi_v$ and $\boldsymbol{u} \cdot \boldsymbol{F}_{st}$, which are obtained by taking the dot product of $\boldsymbol{u}$ and the momentum balance equation. The obtained results are used to verify energy conservation by checking the balance between the energy terms on the left-hand and right-hand sides of Eq. (18). As expected from the application of the divergence theorem, $\rho \boldsymbol{u} \cdot (\boldsymbol{u} \cdot \nabla \boldsymbol{u})$ and $-\boldsymbol{u} \cdot \nabla p$ yield values that are very close to zero. The comparison of the terms on the left-hand side (the time derivative of the kinetic energy) and on the right-hand side (the time derivative of the viscous power plus the work done by the capillary force) is presented in Fig. 6 (b). The figure reveals an excellent balance between the energy terms. This consistency implies that the code accurately maintains momentum conservation with minimal numerical artifacts, while preserving a solenoidal velocity field. The results plotted in Fig. 6 relate to the energy rate and are rendered dimensionless using $\tilde{\tilde{E}}$.

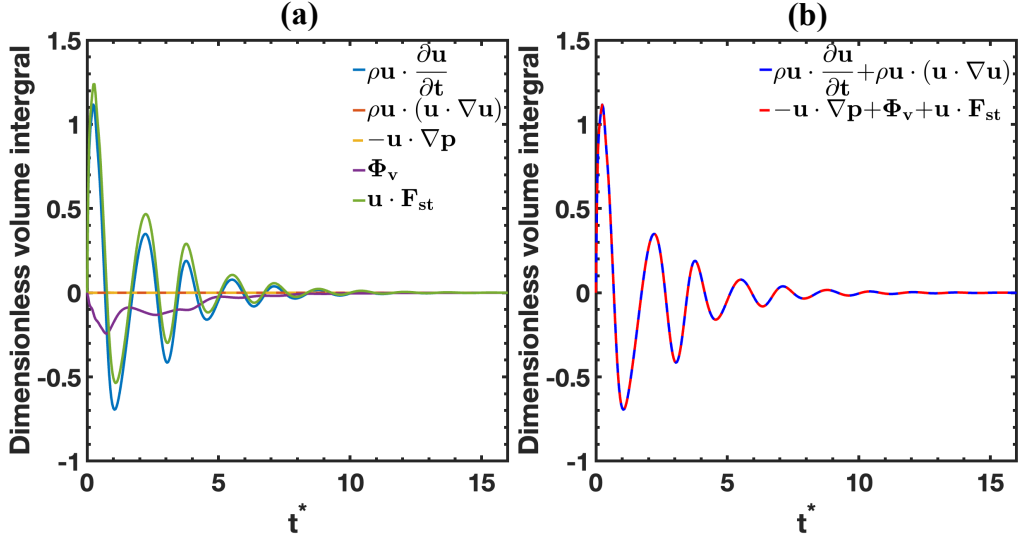

**Fig. 6.** The results for the momentum conservation analysis: (a) the individual contributions from each term in the momentum conservation [Eq. (18)], and (b) the momentum conservation during the simulation.

Having demonstrated that the present numerical framework reliably solves the Navier-Stokes equation while conserving momentum, the effects of the interfacial thickness on different forms of energy are explored. Four different values for the capillary width are considered, such as $\xi = R/10, R/20, R/30$ and $R/40$. First, the effect of $\xi$ on $E_{tot}^*$ is reported. From the energy stability point of view, per Eq. (8), it is expected that $E_{tot}^*$ monotonically decreases in time and $dE_{tot}^*/dt < 0$. The plots in Fig. 7 (a) reveal that the energy dissipation is evident in the results obtained, indicating that the solution of the coupled CHNS equations conserves momentum and obeys the second law of thermodynamics. Once the solution is confirmed to be momentum-conservative and energy stable, the interrelation between the kinetic energy, viscous dissipation, and mixing energy can be accurately assessed using the results plotted in Fig. 7 (b)-(d). All the energy plots imply that the oscillation dynamics are incorrectly predicted when $\xi = R/10$, as evidenced by discrepancies in the peaks of both the kinetic and mixing energies. However, when the interfacial thickness is reduced to below $\xi = R/30$, the energy components exhibit exceptional agreement.

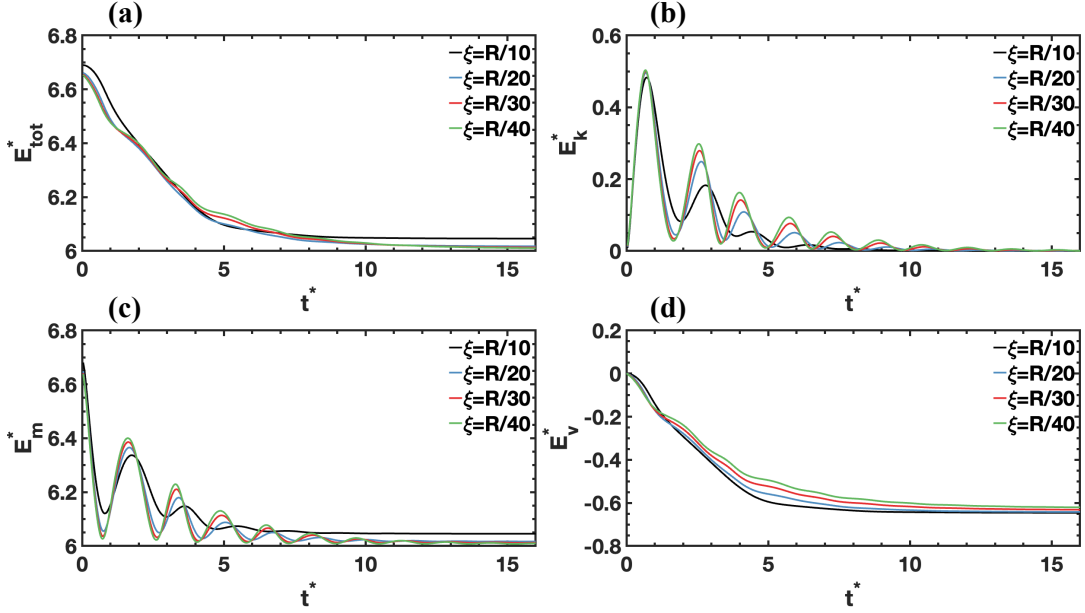

**Fig. 7.** The effect of the interfacial thickness on dimensionless (a) total energy, (b) kinetic energy, (c) mixing energy, and (d) viscous dissipation.

6.3. Unmatched-density droplet oscillation

Rayleigh's linear theory provides an analytical expression for the oscillation frequency of an inviscid droplet in vacuum under infinitesimal-amplitude perturbations [63]. Using the normal-mode technique, the squared eigenfrequency of the $n$th spherical mode (with $n = 2,3, ...$) is

$$\omega_n^2 = n(n-1)(n+2)\frac{\gamma}{\rho R^3}. \tag{42}$$

To validate the present phase-field methodology against the linear theory, oscillations of droplets initially perturbed from a spherical shape are simulated, with the perturbed surface defined by [64]

$$f(\theta) = R_n[1 + \epsilon_0 P_n(\cos\theta)], \tag{43}$$

in which $P_n$ is the Legendre polynomial of order $n$, $\epsilon_0$ is the initial disturbance amplitude and $R_n$ chosen to preserve droplet volume [65]. In these simulations a Reynolds number $Re = 100$ and $\epsilon_0 = 0.01$ are adopted. In Table 1, the numerical oscillation period, $\tau_{PFM}$, is compared with the analytical period $\tau_{analytical} = 2\pi/\omega_n$, from Eq. (42), demonstrating excellent agreement for small-amplitude oscillations.

Table 1. Comparison of oscillation periods between phase-field simulations and linear theory.

| $n$ | $\tau_{PFM}(ms)$ | $\tau_{analytical}(ms)$ | $|\% \, Error|$ |
|---|---|---|---|
| 2 | 0.4132 | 0.4191 | 1.41 |
| 3 | 0.2151 | 0.2164 | 0.60 |
| 4 | 0.1396 | 0.1397 | 0.07 |
| 5 | 0.1005 | 0.1002 | 0.30 |

The extension of the energy analysis to cases with unmatched densities is straightforward. The primary difference arises from the change in the kinetic energy, which is now governed by Eq. (31) instead of Eq. (21), due to the time dependence of the density. The problem is set following the same procedure as described in section 6.2, except for the different values of $Re = 20$, $\rho_1/\rho_2 = 10$ and $\mu_1/\mu_2 = 2$. The present section examines the effect of the interfacial thickness on the total energy and compares between the mixing and surface energies under transient conditions.

The results displayed in Fig. 8 (a) and Fig. 8 (b) reveal that the energy stability is maintained for all the capillary widths. The entire CHNS theory is based on the minimization of the energy functional ($E_{tot}$). When examining $E_{tot}^*$, one finds that the flow is dissipative in nature, primarily due to viscous and diffusive dissipations. Another aspect of the flow, revealed by exemining $E_{tot}^*$, is its equilibrium state. At equilibrium, the fluid velocity approaches zero, eliminating viscous dissipation; however, an unphysical diffusive dissipation may still persist due to the interface regularization term ($\boldsymbol{\nabla} \cdot (M\boldsymbol{\nabla}\psi)$) in the advective Cahn-Hilliard equation. Consequently, numerical simulations can be terminated once viscous dissipation ceases to be significant and $E_{tot}^*$ reaches a plateau. Furthermore, Fig. 8 (a) illustrates the convergence of $E_{tot}^*$ with decreasing $\xi$, particularly as it falls below $R/30$ and approaches the sharp-interface limit inherent to the phase-field formulation.

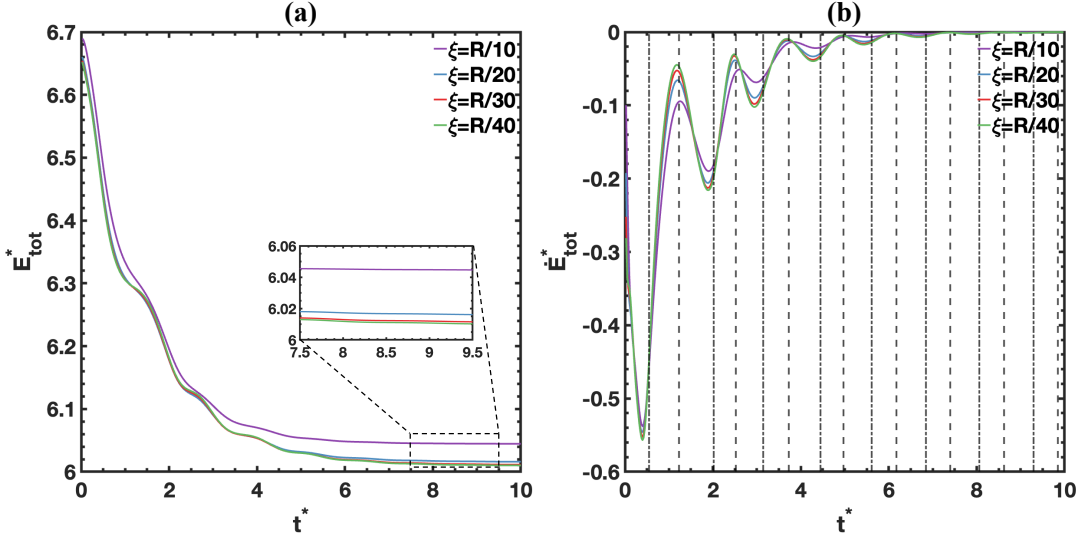

**Fig. 8.** The results for the droplet oscillation with unmatched densities. (a) The time variation of the total energy and (b) the rate of change of the total energy in time. The black dashed lines correspond to the moments with maximum/minimum aspect ratios, while the black dash-dotted lines mark the moments when the aspect ratio equals 1.

On the other hand, examining $\dot{E}^*_{tot}$ provides additional information about the fluid flow. Since it contains the effects of the viscous dissipation [Eq. (8)], it can reveal at which stages the dissipative effects are more pronounced. At the start, with an initial oblate aspect ratio of 2, the droplet possesses an elongated ellipsoidal shape. The surface tension will generate internal flows to reduce the surface area, causing the droplet to deform toward a spherical shape. As the oblate droplet begins to oscillate, the internal flows induced by the capillary forces are relatively strong, resulting in moderate viscous dissipation. During the deformation, as the droplet evolves from an elongated shape toward a spherical one, a substantial internal velocity gradient develops [cf. Fig. 9]. The fluid circulates within the droplet driven by capillary actions, particularly near the interfacial regions where the curvature changes rapidly. Viscous dissipation peaks at intermediate stages, when the droplet's shape is neither fully elongated nor fully spherical. These points correspond to the local minima in Fig. 8 (b), where the droplet undergoes rapid deformation, generating intense internal fluid motion and viscous stresses. As the droplet approaches an ultimate spherical configuration, the internal velocity gradient diminishes, and the fluid motion gradually subsides due to viscous damping. Once the droplet reaches a near-spherical shape, viscous dissipation significantly decreases. At such a near-equilibrium state, the internal fluid flows are minimal, resulting in a minimal dissipation rate. While being always negative, $\dot{E}^*_{tot} \approx 0$ implies the equilibrium stage.

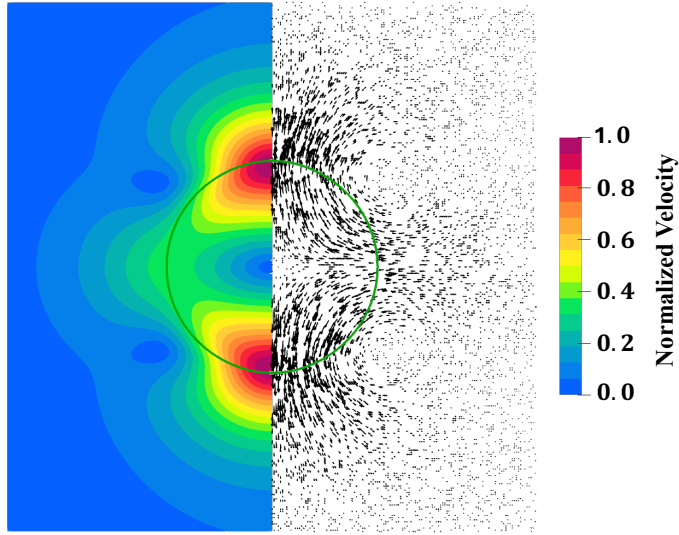

**Fig. 9.** Velocity field around the droplet (outlined in green) at $t^* = 0.54$, corresponding to a unity aspect ratio during the first oscillation period. The left half shows contour plot of the normalized velocity magnitude, and the right half depicts the corresponding velocity vectors in the droplet's vicinity.

The total energy $E_{tot}$ is obtained via Eq. (9) and encompasses both kinetic and mixing energies. Intuitively, the total energy of the two-phase flow comprises both kinetic and surface energies. As mentioned before, the relationship between mixing and surface energies is defined in Eq. (14) under steady-state conditions. However, here a detailed analysis of the correlation between the two types of energy is pursued under transient conditions. To calculate the surface energy, the interface profile is extracted as the isocontour of $c = 0$ with ParaView [66]. The integration in Eq. (39) requires the interfacial profile data to be sorted for evaluating the surface area. A nearest neighbor algorithm is utilized to achieve the necessary sorting of the interface profile data. The mixing energy $E_m$ or the free energy functional $\mathcal{F}$ composed of the bulk and gradient components is obtained by evaluating the volume integral in Eq. (7).

The results plotted in Fig. 10 (a) and Fig. 10 (b) reveal that reducing the interfacial thickness facilitates reduction of the difference between the mixing and surface energies, consistent with the static tests. While transient simulations capture the surface energy with high fidelity, they exhibit marginally higher errors than the static tests. This increase stems from three principal deviations from the assumptions underlying Eq. (14). First, the fourth-order spatial derivative in the Cahn-Hilliard equation produces minor undershoots and overshoots in $c$. These oscillations perturb the integral in Eq. (7). Second, the interface thickness—assumed constant normal to the interface in the static tests—can vary during transient evolution [67,68]. Third, the numerical solution for $c$ departs subtly from the analytical hyperbolic-tangent form of Eq. (12). Together, these factors explain the elevated error observed under transient conditions.

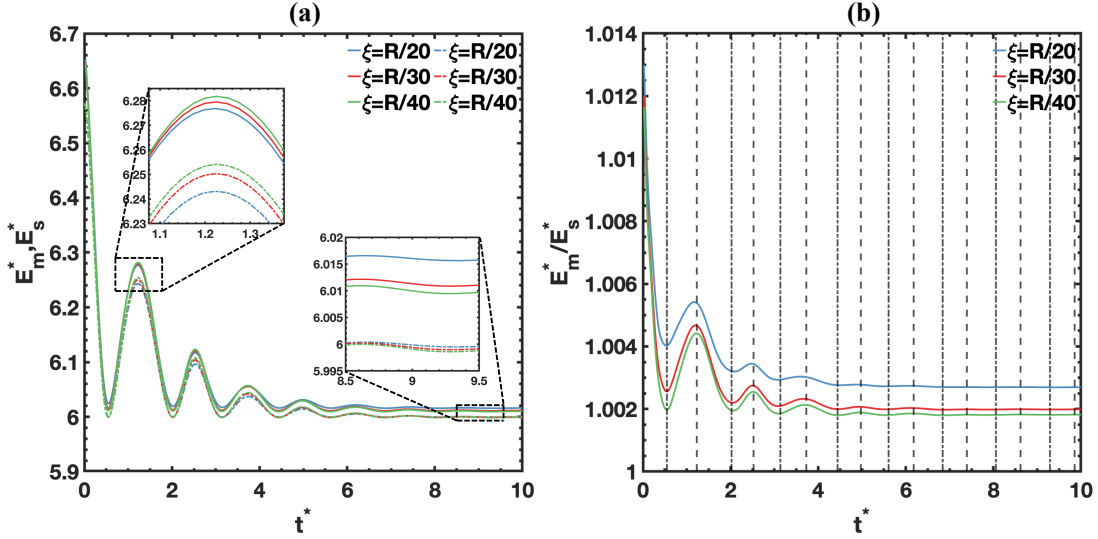

**Fig. 10.** Comparison between the mixing energy and the ordinary surface energy. (a) Dimensionless evolution of the mixing energy (solid lines) and the surface energy (dash-dotted lines) as a function of the non-dimensionalized time $t^*$ and (b) ratio of the mixing energy to the surface energy as a function of $t^*$, with different curves corresponding to various capillary widths $\xi$. The black dashed lines indicate the moments of maximum deformation, while the black dash-dotted lines denote the moments when the aspect ratio equals unity.

Another feature of Fig. 10 is the convergence behavior of $E_m$ and $E_s$. The plots demonstrate excellent convergence with respect to interfacial thickness. In particular, the convergence of $E_s$ with capillary width indicates that the sharp interface (i.e., the $c = 0$ isocontour) remains virtually unchanged across different capillary widths. Moreover, the zoomed-in insets in Fig. 10 (a) reveal that $E_s$ is less affected by numerical integration artifacts than $E_m$, leading to a slightly more robust convergence of the surface energy.

6.4. Breakup of a cylindrical liquid thread

The breakup of a liquid thread is investigated using the diffuse interface approach. The temporal instability of an axisymmetric capillary thread is numerically simulated, triggered by an initial harmonic perturbation. The simulation protocol and boundary conditions follow the procedure established by Ashgriz and Mashayek [69]. The phase-field variable initialization is given by the following hyperbolic tangent function in a square domain $[0, \Lambda] \times [0, \Lambda]$:

$$c(r, z) = \tanh \frac{2\left(R + \varepsilon_0 R \cos\left(\frac{2\pi z}{\Lambda}\right) - r\right)}{2\sqrt{2}\xi}, \tag{44}$$

where $R$ is the radius of the undisturbed thread; $\varepsilon_0$ is the amplitude of the initial surface perturbation; $\Lambda$ is the wavelength; $\xi$ is the interfacial thickness, and $k = 2\pi R/\Lambda$ is the wavenumber. The Reynolds and Weber numbers, defined as in the droplet-oscillation study, are set to $Re = 10$ and $We = 1$, and the initial perturbation amplitude is $\epsilon_0 = 0.05$.

The numerical code is validated by comparing the droplet radii predicted by the phase-field method with the experimental results of Lafrance [70]. In Fig. 11, the numerical and experimental radii of both primary and satellite droplets exhibit reasonable agreement over the wavenumber range $k = 0.2 - 0.9$.

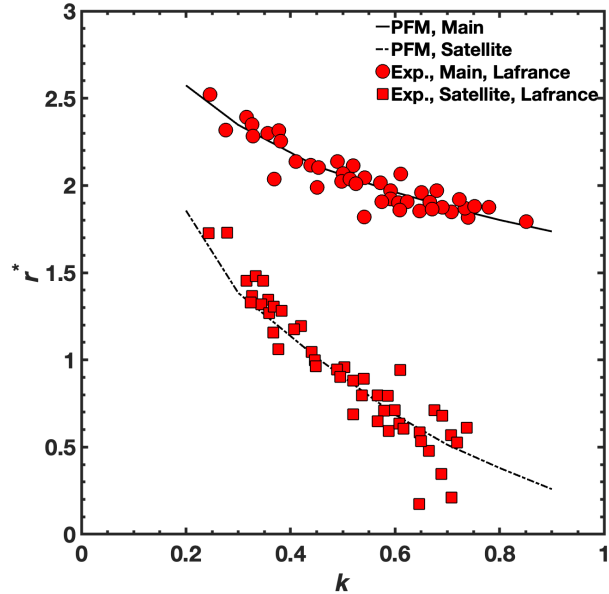

**Fig. 11**. Comparison of the main and satellite droplet radii predicted by the present phase-field model (PFM) with the experimental results of Lafrance [70], plotted as a function of wavenumber.

For the energy analysis, the wavenumber is fixed at $k = 0.5$. The density and viscosity ratios are selected as, $\rho_1/\rho_2 = 100$ and $\mu_1/\mu_2 = 10$, respectively. The future analysis aims at examining the dynamics of energy transfer at the moment of breakup and during the oscillations of the main and satellite droplets after breakup. It also quantifies the kinetic energy in both droplets post-pinch-off to confirm the presence of severely varying scales of motion. Finally, it seeks to compare the work of the surface tension force with the variations in the mixing and surface energies to assess the accuracy of the present capillary force formulation.

In the framework of the numerical simulations of capillary breakup, the interplay between the kinetic and surface energy, and viscous dissipation during droplet breakup has been scarcely explored. Most phase-field-based numerical simulations in the two-phase flow literature focus on the primary jet/thread breakup, while satellite droplet formation has long been the subject of experimental studies, particularly in inkjet printing [71]. Due to the robustness of the energy-based phase-field methods, they do not require special attention to complex topological changes such as droplet breakup or coalescence. Despite its advantages, the Cahn-Hilliard model is well known to suffer from an unphysical mass loss [72]. This issue is exacerbated when small and large features coexist, as in the case of main and satellite droplets in liquid-thread breakup. It can lead to the eventual disappearance of smaller droplets or bubbles through a process known as Ostwald ripening, or coarsening [73,74]. See [75,76] for strategies to mitigate Ostwald ripening, including modifications to the free energy functional and the use of degenerate mobility formulations. In this work, to prevent the unphysical disappearance of satellite droplets with smaller radii, an adaptive time stepper is employed to adjust the time step size to match the characteristic oscillation time scale of the satellite droplets after pinch-off [46]. Employing this technique, the unphysical shrinkage of smaller droplets is effectively prevented [46]. The initial profile of the capillary thread and some results before the primary breakup are given in Fig. 12 (a) and post-breakup in Fig. 12 (b).

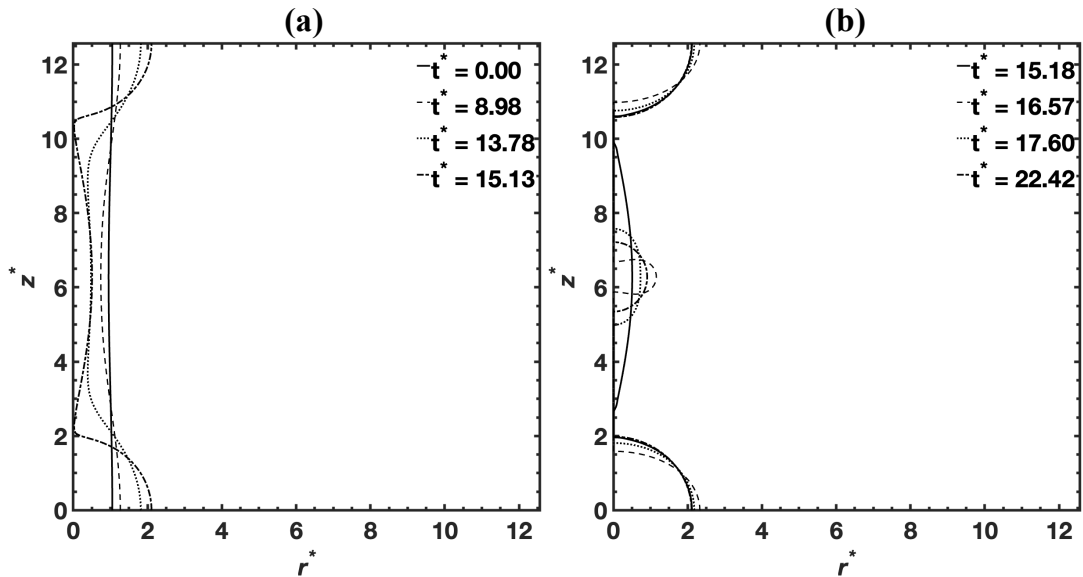

**Fig. 12.** The results for the breakup of a capillary thread. Time sequence of the thread evolution for (a) primary breakup and (b) post-breakup droplet oscillations. The perturbation is periodic along the thread.

Initially, the thread possesses maximum surface or mixing energy and zero kinetic energy, as in Fig. 13 (a) and Fig. 13 (b). Upon the perturbation growth, the kinetic energy of the flow continuously increases until $t^* = 13.78$ and facilitates the linear and nonlinear evolution of the initial harmonic perturbation. Close to this moment, the thread rapidly evolves toward the pinch-off point, and the flow accelerates due to the capillary forces driving it from the thinner neck zone into the thicker swell regions. Between $t^* = 13.78$ and $t^* = 15.04$, as the flow evolves toward the pinch-off, high local viscous dissipation in high-curvature regions drains the kinetic energy of the two-phase flow.

At about $t^* = 15.13$, a singular point forms on the thread where the radius approaches zero and the surface curvature increases dramatically. This is roughly the moment when breakup occurs, accompanied by a significant increase in the kinetic energy. In addition to $E_k^*$ and $E_m^*$, the time derivatives of the kinetic and the mixing energies ($\dot{E}_k^*$ and $\dot{E}_m^*$) provide a significant physical insight into the thread breakup. Fig. 13 (a) and Fig. 13 (c) reveal that as soon as the thread breakup happens, the kinetic energy of the flow ($E_k^*$) substantially increases. The maximum liquid velocity occurs near the pinch-off location at the breakup moment, which accelerates the satellite droplet after the primary breakup.

At the pinch-off moment, the time derivative of the mixing energy ($\dot{E}_m^*$) reaches its minimum value, meaning that the rate of the decrease in the mixing/surface energy reaches its highest magnitude. Therefore, it is $\dot{E}_m^*$ not $E_m^*$, that exhibits extrema at the breakup moment [cf. Fig. 13 (b) and Fig. 13 (d)]. Then, $E_m^*$ continues to decrease until the dynamics become dominated by the oscillations of the main and satellite droplets.

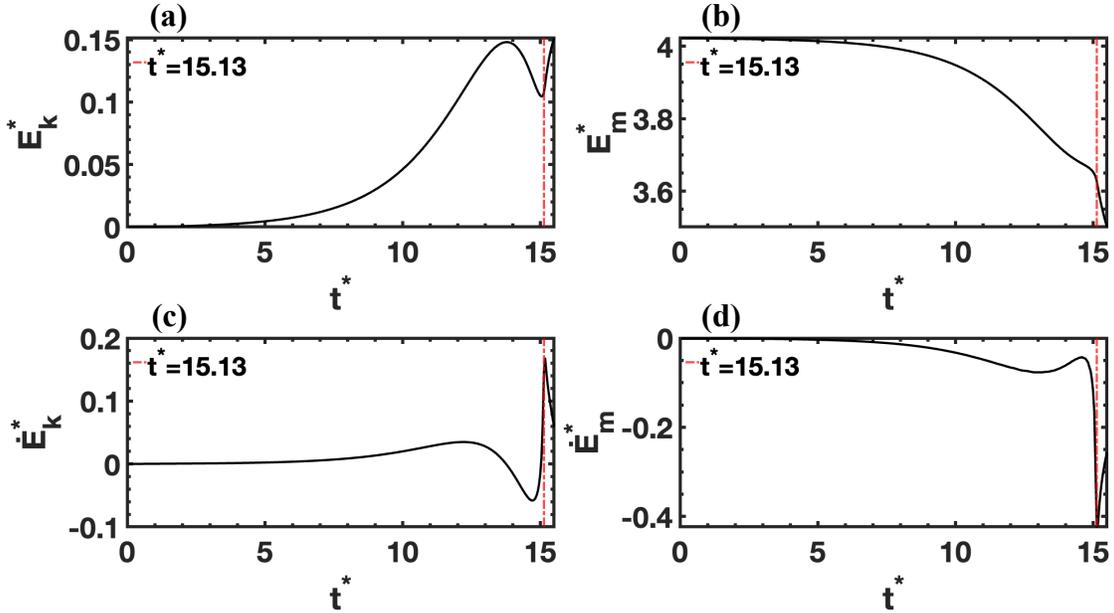

**Fig. 13.** The results for the temporal evolution of the (a) kinetic energy, (b) the mixing energy, (c) the time derivative of the kinetic energy, and (d) the time derivative of the mixing energy during the capillary thread breakup.

The non-dimensional contour plots of the kinetic energy density $e_k = \frac{1}{2}\rho|\boldsymbol{u}|^2$ and viscous dissipation $\phi_v = -\boldsymbol{\tau}:\nabla\boldsymbol{u}$ are shown in Fig. 14, with the sharp fluid–fluid interface outlined by the curvature magnitude $\kappa$ normalized by $1/R$. The field $e_k^*$ highlights regions of high velocity magnitude, while $\phi_v^*$ shows zones of intense velocity gradients and viscous loss. The viscous dissipation patterns indicate that viscous loss is greatest in regions with high curvature gradients and rapid shear rates. In particular, $\phi_v^*$ peaks in the neck region where curvature gradients—and hence $\nabla\boldsymbol{u}$—are largest. As the droplet evolves between $t^* = 12.14$ and $t^* = 14.78$, neck thinning intensifies these curvature gradients and causes an explosive increase in the maximum $\phi_v^*$, signaling imminent pinch-off.

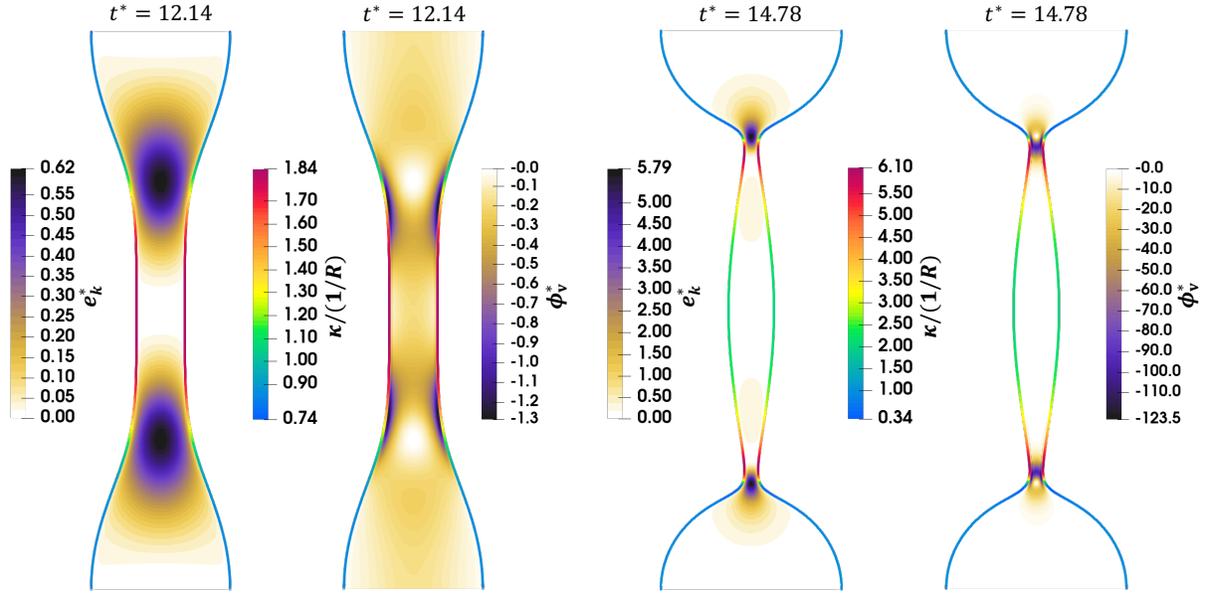

**Fig. 14.** Contours of the non-dimensional kinetic energy density ($e_k^*$) and viscous dissipation ($\phi_v^*$) inside the liquid thread pre-breakup at $t^* = 12.14, 14.78$. The sharp interface is outlined by curvature magnitude $\kappa$ normalized by $1/R$.

Post-breakup, a smaller satellite droplet with a high velocity inside and a larger main droplet with a lower velocity inside emerge. While the physics of the flow prior to the primary breakup is well understood, the post-breakup behavior, particularly concerning satellite droplet(s), has been less explored. Despite the advancements in experimental studies, numerical modeling of post-breakup dynamics is still not well developed.

Droplet oscillation is heavily influenced by its size and initial perturbation [63–65,77,78]. The results presented in Fig. 15 illustrate the shape of the satellite droplet at various stages. The first oscillation period measured as the time between the first and second maximum axial deformations differs from the subsequent periods. This stems from the highly deformed droplet shape after the pinch-off, and from the high value of $\dot{E}_k^*$ at the breakup time. At $t^* = 15.18$, the ligament is long and thin, indicating that it has been stretched significantly. By $t^* = 15.71$, the droplet shape has become more constricted, with a visible neck forming in the middle. This indicates that the ligament is relaxing due to the surface tension but is not undergoing further breakup. Between $t^* = 15.71$ and $t^* = 16.61$, the ligament undergoes a remarkable transformation. As time progresses, the middle section narrows, and the ligament begins to evolve toward a more rounded form. By $t^* = 16.61$, the droplet shape adopts a distinct toroidal (a donut-like) configuration, indicating a transitional state driven by the surface tension. This toroidal droplet shape signifies the retraction of the fluid body, as it moves toward a stable configuration before ultimately evolving into a spherical droplet. After the first oscillation period, the satellite droplet undergoes regular transitions between oblate and prolate spheroidal shapes. The oscillations continue until the droplet quickly stabilizes into its final equilibrium spherical form.

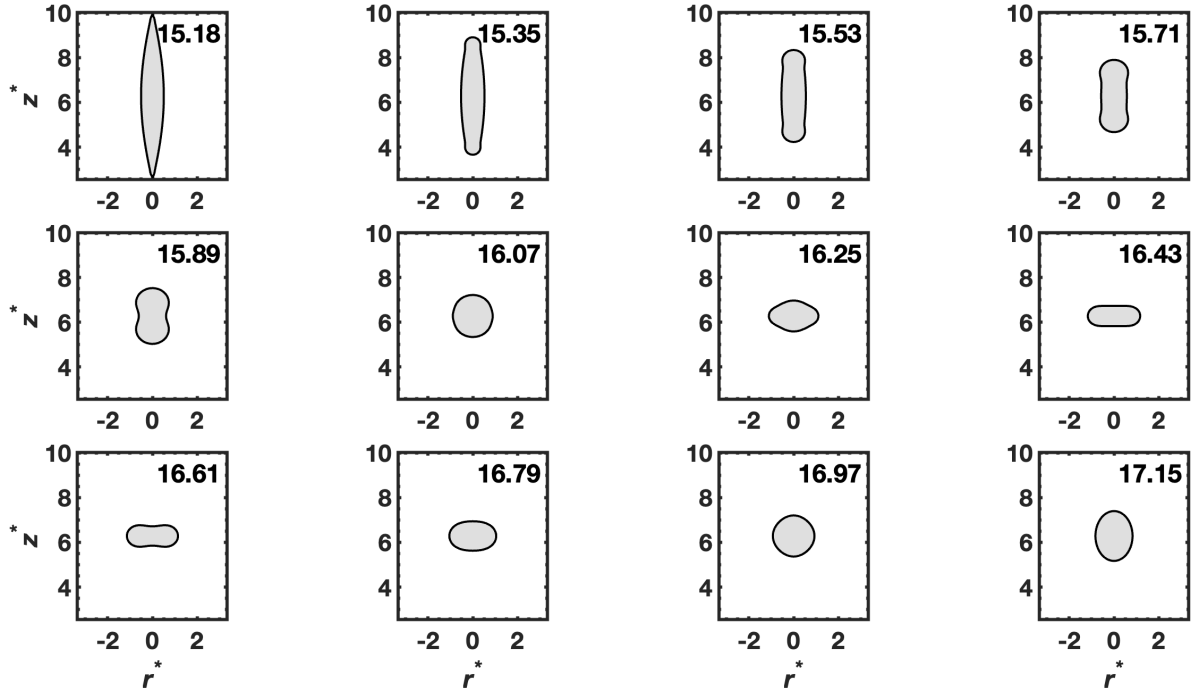

**Fig. 15.** Time sequence of the oscillations of a satellite droplet after pinch-off. The numbers in the figures represent the non-dimensional time.

In Fig. 16, the dimensionless velocity magnitude within the droplet and the flow streamlines in the outer fluid are depicted. In panels (a) and (b), the thinning of the thread is evident, with high-velocity regions concentrated near the pinch-off location. Immediately after breakup in panel (c), the velocity magnitude peaks at the satellite droplet tip, rapidly driving its motion. In panel (d), the lower velocity of the main droplet compared to the satellite droplet is observable. The velocity field is rendered dimensionless using the maximum velocity at the tip of the satellite droplet post-pinch-off at $t^* = 15.18$. Moreover, the outer fluid streamlines further illustrate the interaction between the droplet and the surrounding medium, highlighting the recirculation zones that dynamically evolve before and after breakup in response to the droplet deformation.

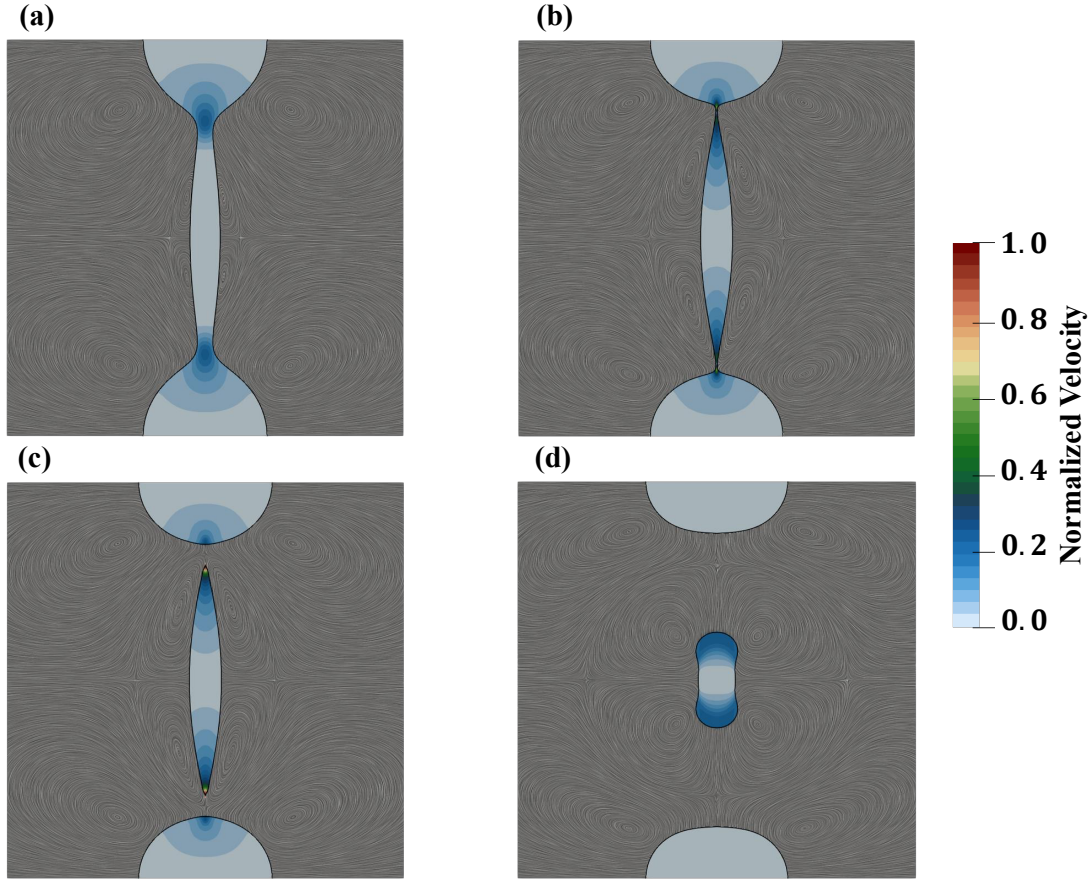

**Fig. 16.** Visualization of the normalized velocity magnitude inside the droplets and the flow streamlines in the surrounding fluid before and after thread breakup, shown sequentially. The results are illustrated as follows: (a) $t^* = 14.50$, (b) $t^* = 15.13$, (c) $t^* = 15.18$, and (d) $t^* = 15.76$.

The total kinetic energies of the main and satellite droplets displayed in Fig. 17 confirm the existence of markedly different temporal scales of motion between the main and satellite droplets. Before the breakup, a high curvature at the pinch-off location generates a significant pressure gradient, accelerating the liquid and resulting in a high velocity at the satellite droplet's tip [cf. Fig. 16 (b) and Fig. 16 (c)]. Consequently, the differences in fluid velocity, along with the size and shape disparities, contribute to the distinct time scales of motion between the satellite droplet and the main one. Both droplets attempt to evolve toward a spherical shape, minimizing their surface area and energy. Right after the primary breakup, the satellite droplet accounts for nearly 28% of the total kinetic energy, while carrying only 8% of the total mass of the original thread section corresponding to one perturbation wavelength. The figure indicates that the first oscillation of the satellite droplet is accompanied by the strongest increase in the kinetic energy, corresponding to a decrease in the surface energy. At $t^* = 15.98$ where the satellite droplet acquires its maximum kinetic energy, it accounts for 88% of the total kinetic energy. Yet, the smaller size of the satellite droplet results in a lower Reynolds number, leading to faster oscillations. Its oscillations dampen more rapidly than those of the main droplet due to stronger viscous effects. The results for the kinetic energy indicate that all periods of the oscillations of the satellite droplet are smaller than those for the main droplet. After five oscillations, the satellite droplet ceases its oscillations and reaches an equilibrium state by $t^* = 22$. On the other hand, the main droplet continues to oscillate, with its motion persisting even beyond $t^* = 36$, retaining all the kinetic energy from the pinched-off thread.

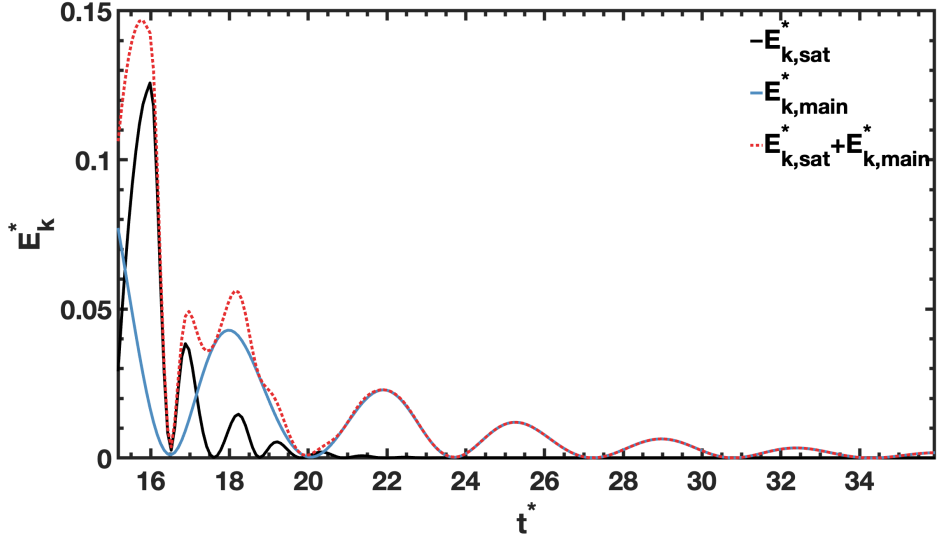

**Fig. 17.** Temporal variation of the kinetic energy of the main and satellite droplets, along with their combined total kinetic energy.

The final objective of the present study is to provide the value of the integral $\int_0^t \int_\Omega \boldsymbol{u} \cdot \boldsymbol{F}_{st}\, d\Omega dt$, which essentially indicates the work of the capillary force, and reveal the comparison with the changes in the mixing and surface energies. This allows one to quantitatively assess the accuracy of the capillary force derived from the Korteweg stress tensor [Eq. (A1)] in representing the changes in the surface/mixing energy. The results displayed in Fig. 18 compare the dimensionless integral $\int_0^t \int_\Omega \boldsymbol{u} \cdot \boldsymbol{F}_{st}\, d\Omega dt$, with $\Delta E_m^*$ and $\Delta E_s^*$. At the initial stage when the thread is almost smooth with a small harmonic perturbation, these variables remain indistinguishable. As the thread develops and the curvature changes across the thread interface, the disparity between $\Delta E_m^*$ and $\Delta E_s^*$ increases. The reasons for this discrepancy are detailed in section 6.3, where $E_m^*$ and $E_s^*$ are compared under transient conditions. Despite this, by providing a sufficient mesh resolution using AMR, the work of the capillary force and $\Delta E_m^*$ closely match throughout the entire computational period, further validating the accuracy of the present surface tension formulation.

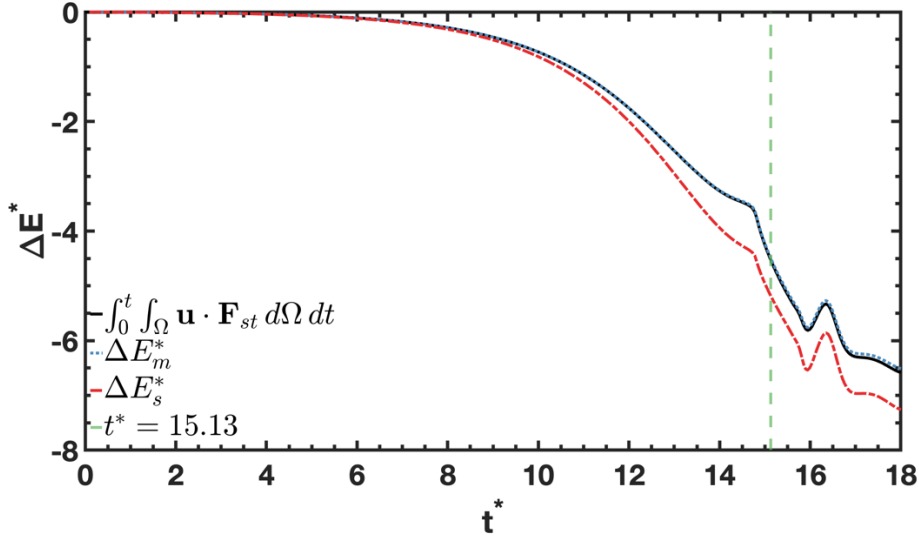

**Fig. 18.** Comparison between the dimensionless work of the capillary force and the change in the mixing energy and the surface energy for $\xi = R/30$. The green dashed line corresponds to the breakup moment.

## 7. Conclusions

The present work provides a comprehensive energy-based perspective on two-phase flows modeled using the coupled CHNS equations. By ensuring momentum conservation in the numerical framework, the critical role of the energy balance is highlighted for the accurate prediction of the two-phase flow dynamics. The energy equation is fundamental for this analysis, offering insights into the interplay between the kinetic energy, mixing energy, and viscous dissipation.

A central focus of the present study is the comparison between the mixing energy and the surface energy. Static tests demonstrate that the mixing-energy theory provides highly accurate surface energy approximations for curved interfaces with sharp curvature gradients. However, in transient simulations, undershoots and overshoots, non-uniform interface thickness, and departures from the equilibrium hyperbolic-tangent profile increase errors in the surface energy estimates. The energy-based surface tension formulation, derived from the Korteweg stress tensor, effectively represents variations in the mixing energy, further validating its robustness for modelling complex interfacial flows.

The dynamics of droplet oscillations and capillary thread breakup are investigated as benchmark tests, revealing new insights into two-phase flow dynamics. For droplet oscillations, the findings confirm that reducing the interfacial thickness enhances convergence of the energy components, enabling an accurate characterization of the interaction between the kinetic energy, viscous dissipation, and mixing energy. Furthermore, the simulations identify the temporal intervals during which viscous dissipation dominates the energy dynamics. For capillary thread breakup, the analysis highlights distinct energy transfer mechanisms at the moment of breakup, specifically the rise in the kinetic energy and a simultaneous abrupt decrease in the mixing energy. Post-breakup, the satellite droplet dynamics exhibit faster oscillation damping due to its lower Reynolds number, contrasting with the persistent oscillations of the main droplet.

The time derivative of the total energy functional emerges as a powerful tool for identifying dissipative flow stages, providing detailed information regarding the energy transfer mechanisms. The present work not only examines two-phase flows from an energy standpoint but also offers practical methodologies to ensure numerical consistency and accuracy in terms of the momentum conservation. A future work should extend the energy-based approach to incorporate additional physical phenomena, e.g., gravity, Marangoni flows [79], thermal effects, and the effect of the electric field.

**CRediT authorship contribution statement**

**Ali Mostafavi:** Conceptualization (lead); Formal analysis (lead); Methodology (lead); Visualization (lead); Writing – original draft (lead). **Mohammadmahdi Ranjbar:** Conceptualization (supporting); Formal analysis (supporting), Writing – review & editing (equal). **Vitaliy Yurkiv**: Conceptualization (supporting); Writing – review & editing (equal). **Alexander L. Yarin:** Conceptualization (supporting); Methodology (supporting); Writing – review & editing (equal). **Farzad Mashayek**: Funding acquisition (lead); Supervision (lead); Methodology (supporting); Formal analysis (supporting), Writing – review & editing (lead).

**Declaration of competing interest**

The authors declare that they have no known competing financial interests or personal relationships that could have appeared to influence the work reported in this paper.

**Data availability**

Data will be made available on request.

**Acknowledgements**

The authors acknowledge the financial support from the National Science Foundation awards CBET 2312197 and 2224749. Simulations were performed using the High-Performance Computing (HPC) resources supported by the University of Arizona TRIF, UITS, and Research, Innovation, and Impact (RII) and maintained by the University of Arizona Research Technologies team.

**Appendix: Derivation of the capillary force from the Korteweg stress tensor**

In the phase-field methods, the capillary force is calculated through the second law of thermodynamics [34,47]. This force is proportional to the divergence of the so-called Korteweg stress tensor $\nabla \cdot (\nabla c \otimes \nabla c)$ where $\otimes$ is the tensor (dyadic) product which satisfies the energy dissipative law [80] and is obtained as follows [34]:

$$\sigma \nabla \cdot (\nabla c \otimes \nabla c) = \sigma(\nabla c \nabla^2 c + \nabla c \cdot \nabla(\nabla c)) = \sigma\left(\nabla c \nabla^2 c + \frac{1}{2}\nabla(\nabla c \cdot \nabla c)\right). \tag{A1}$$

The above equation can be modified such that the chemical potential, $\psi$, appears by adding and subtracting $f'\nabla c$ where $f' = \partial f(c)/\partial c$:

$$\sigma \nabla \cdot (\nabla c \otimes \nabla c) = \sigma\left(\nabla c \nabla^2 c + \frac{1}{2}\nabla(\nabla c \cdot \nabla c) + f'\nabla c - f'\nabla c\right). \tag{A2}$$

A further simplification invokes Eq. (4):

$$\sigma \nabla \cdot (\nabla c \otimes \nabla c) = \sigma \nabla c(\nabla^2 c - f') + \sigma\left(\frac{1}{2}\nabla(\nabla c \cdot \nabla c) + f'\nabla c\right), \tag{A3}$$

where $\psi = \sigma(f' - \nabla^2 c)$. Putting the second term on the right-hand side in Eq. (A3) inside the $\nabla$ operator results in:

$$\sigma \nabla \cdot (\nabla c \otimes \nabla c) = -\psi \nabla c + \sigma \nabla\left(\frac{1}{2}\nabla c \cdot \nabla c + f(c)\right). \tag{A4}$$

Using the Korteweg stress tensor on the right-hand side of the momentum balance equation changes the tensor sign. The second term on the right-hand side in Eq. (A4) can be absorbed in the fluid pressure gradient term $\nabla P$. Therefore, the following approximate expressions are obtained for the volumetric capillary force and modified pressure:

$$\boldsymbol{F}_{st} = \psi \nabla c, \tag{A5}$$

$$p = P + \sigma\left(\frac{1}{2}\nabla c \cdot \nabla c + f(c)\right). \tag{A6}$$

Consequently, the energy-based capillary force is derived, and the fluid pressure $P$ in Eq. (2) is replaced by the modified pressure $p$. All equations remain invariant and thus, are applicable in 2D, 2D axisymmetric, and 3D contexts.